\documentclass[journal,article,submit,pdftex,moreauthors]{mdpi} 

\firstpage{1} 
\pubvolume{1}
\issuenum{1}
\articlenumber{0}
\pubyear{2023}
\copyrightyear{2023}
\datereceived{ } 
\daterevised{ } 
\dateaccepted{ } 
\datepublished{ } 
\hreflink{https://doi.org/} 

\Title{Pion interferometry with L\'evy-stable sources in $\sqrt{s_{NN}}$~=~200~GeV Au+Au collisions at STAR}
\TitleCitation{Pion interferometry with L\'evy-stable sources in $\sqrt{s_{NN}}$~=~200~GeV Au+Au collisions at STAR}

\Author{D\'aniel Kincses $^{1}$\orcidA{} for the STAR Collaboration}

\AuthorNames{D\'aniel Kincses}

\AuthorCitation{Kincses, D.}

\address{%
$^{1}$ \quad ELTE E\"otv\"os Lor\'and University, H-1117 Budapest, P\'azm\'any P\'eter s\'et\'any 1/A; kincses@ttk.elte.hu}

\abstract{Measurements of femtoscopic correlations in high-energy heavy-ion collisions aim to unravel the space-time structure of the particle-emitting source (the quark-gluon plasma). Recent results indicate that the pion pair source exhibits a power-law behavior and can be described well by a L\'evy distribution. In this study, L\'evy fits were performed to the measured one-dimensional two-pion correlation functions in Au+Au collisions at $\sqrt{s_{NN}}$=200~GeV. The three extracted source parameters are the L\'evy scale parameter, $R$, which relates to the size of the source; the correlation strength parameter, $\lambda$; and the L\'evy exponent, $\alpha$, which characterizes the power-law tail of the source. In this paper, we report the current status of the analysis of the extracted L\'evy source parameters and present their dependence on average transverse mass, $m_T$, and on centrality. 
}


\begin{document}

\section{Introduction to femtoscopic correlations}

One of the indispensable tools aiding the quest to explore the matter created in high-energy collisions of heavy nuclei is femtoscopy~\cite{Boal:1990yh,Weiner:1999th,Wiedemann:1999qn,Csorgo:1999sj,Kisiel:2011jt,Lisa:2005dd}. Femtoscopy utilizes quantum statistics and final-state interactions to make a connection between momentum correlations and spatial correlations. Its name was coined because, with the help of such correlation measurements, one can map the space-time geometry of the particle-emitting source on the femtometer scale. In the following, we review some of the basic definitions of femtoscopic correlation functions and discuss the shape of the two-particle source. Although femtoscopic techniques can be applied to a large variety of particle combinations~\cite{Lednicky:2001qv,Chakraborty:2022jpf}, in this paper, we focus on two-particle correlations of identical pions, and thus, we introduce the following definitions accordingly.

Utilizing the $q = p_1 - p_2$ relative pair momentum and the $K = 0.5(p_1+p_2)$ average pair momentum variables, the two-particle momentum correlation function $C_2(q,K)$ can be expressed with the pair source function $D(r,K)$ and the symmetrized pair wave function $\psi_q(r)$ as

\begin{align}
    C_2(q,K) = \int d^4r D(r,K)|\psi_q(r)|^2.
    \label{e:Cq}
\end{align}

The pair wave function contains effects from quantum statistics and final-state interactions and can be calculated from the well-known quantum-mechanical Coulomb problem~\cite{Landau:1991wop}. The pair source function $D(r,K)$ is defined as the auto-correlation of the $S(x,K)$ single-particle source function or phase-space density (where $r$ is the relative coordinate of the pair, and the smoothness approximation $p_1\approx p_2\approx K$ is utilized~\cite{Pratt:1997pw}):

\begin{align}
    D(r,K) = \int d^4\rho\ S\left(\rho+\frac{1}{2}r,K\right)S\left(\rho-\frac{1}{2}r,K\right).
    \label{e:Dr}
\end{align}

The pair source $D(r,K)$ cannot be measured directly in experiments. However, utilizing Eq.~\eqref{e:Cq}, measurements of the momentum correlation function can provide information about the shape of the pair source. The latter, especially for pions, has been under the femtoscope for a long time. A commonly used assumption for the shape of the pion pair source was the Gaussian, or normal distribution. However, there have been indications of a power-law behavior~\cite{Brown:1997ku, PHENIX:2006nml}, and recent studies both in phenomenology~\cite{Kincses:2022eqq,Korodi:2022ohn,Nagy:2023zbg,Kurgyis:2020vbz} and experiment~\cite{PHENIX:2017ino,CMS:2023xyd,NA61SHINE:2023qzr} showed that the more general L\'evy-stable distribution might provide a better description. In this paper, we present the latest preliminary results of L\'evy source parameter measurements at the STAR experiment.

\section{L\'evy-stable source distributions}

As mentioned previously, observations of a heavy tail in the pion source required a more general approach and a need to go beyond the Gaussian description. L\'evy-stable distributions~\cite{Nolan:Levy,Csorgo:2003uv} (arising from the Generalized Central Limit Theorem~\cite{Gnedenko:GCLT}) in case of spherical symmetry are defined as

\begin{align}
    \mathcal{L}(\alpha,R;\boldsymbol{r})=\frac{1}{(2\pi)^3}\int d^3\boldsymbol{q}e^{i\boldsymbol{qr}}e^{-\frac{1}{2}|\boldsymbol{q}R|^\alpha},
\end{align}
where $\alpha$ is the L\'evy exponent parameter, describing the tail of the distribution, and $R$ is the L\'evy scale parameter. Important properties of these stable distributions are that they exhibit a power-law behavior for $\alpha < 2$, and they retain the same $\alpha$ value under convolution, i.e., if the single particle source is a L\'evy distribution, the pair source will be one as well, with the same exponent.

Such distributions were found to be good candidates to describe experimental measurements~\cite{PHENIX:2017ino,CMS:2023xyd,NA61SHINE:2023qzr}, as well as the source shape in event generator models such as EPOS~\cite{Kincses:2022eqq,Korodi:2022ohn}. The reason for the appearance of the apparent $\alpha < 2$ power-law behavior is yet to be understood in detail; at different collision energies and different collision systems, there can be many various competing phenomena, such as jet fragmentation~\cite{Csorgo:2004sr,L3:2011kzb}, critical behavior~\cite{Csorgo:2005it}, event averaging~\cite{Tomasik:2019tjj,Cimerman:2019tku}, resonance decays~\cite{Csanad:2007fr,Kincses:2022eqq,Korodi:2022ohn} and anomalous diffusion~\cite{Metzler:1999zz,Csanad:2007fr}. 

In the case of the top RHIC energy investigated in the current paper, anomalous diffusion (in terms of hadronic rescattering) and resonance decays are the most probable. It was shown in Refs.~\cite{Kincses:2022eqq,Korodi:2022ohn} that in heavy-ion collisions generated by EPOS, the latter cannot be the sole reason for the L\'evy shape; primordial pions after UrQMD also exhibit a power-law behavior. It is also important to note, that these analyses were done on an event-by-event basis, and L\'evy shape appeared in individual events. Thus, (at least in EPOS) it is not the event averaging that is behind the appearance of the L\'evy shape.

Another usual argument in the case of one-dimensional analyses is that angle-averaging leads to a non-Gaussian behavior; this is, of course, true, however, an angle-averaged non-spherical Gaussian is strikingly different from a L\'evy-distribution as it was shown in the talk of M. Csan\'ad at the 52nd International Symposium on Multiparticle Dynamics and the XVI. Workshop of Particle Correlations and Femtoscopy. The details of those talks are also summarized in the same special issue as this paper~\cite{Csanad:2024hva}. In Ref.~\cite{Kurgyis:2018zck}, it was also shown that an angle-averaged experimental L\'evy analysis leads to the same $\alpha$ exponent as a three-dimensional measurement. The angle averaging for L\'evy sources was also investigated in Ref.~\cite{Kurgyis:2020vbz}.

It is easy to show, that in the absence of any final-state interactions, the correlation function is in direct connection with the Fourier-transform of the pair source function~\cite{Nagy:2023zbg}. Thus for spherically symmetric L\'evy sources, it takes the following simple form:

\begin{align}
    C_2(Q) = 1 + \lambda e^{-|RQ|^{\alpha}},
    \label{e:c2levy}
\end{align}
where $Q$ is the magnitude of the relative-momentum variable. In Eq.~\eqref{e:c2levy}, the $\lambda$ correlation strength parameter was also introduced to account for a possible change from unity (caused by, e.g., decays of long-lived resonances~\cite{Csorgo:1994in}, partial coherence~\cite{Csorgo:1999sj}, or quasi-random electromagnetic fields~\cite{Csanad:2020qtu}). The average momentum $K$ dependence in Eq.~\eqref{e:c2levy} appears through the source parameters $\lambda(K)$, $R(K)$, $\alpha(K)$. In experimental measurements, usually, instead of $K$ the average transverse mass is used, defined as ${m_T = \sqrt{k_T^2+m^2}}$, where $m$ is the particle mass and $k_T$ is the transverse component of $K$.

In the case of identical charged pions (which are the subject of the current paper), one cannot neglect the Coulomb final-state interaction. Calculating the shape of the correlation function for L\'evy-stable sources with the Coulomb-interacting pair wave function is not a trivial task, however, numerical calculations are possible, as detailed in Ref.~\cite{Nagy:2023zbg}. Taking into account the Coulomb-interaction, Eq.~\eqref{e:c2levy} can be modified to the following:

\begin{align}
    C_2(Q) = \left(1-\lambda+\lambda\cdot K(Q;\alpha,R)\cdot\left(1+e^{-|RQ|^\alpha}\right)\right)\cdot N(1+\varepsilon Q),
    \label{e:fitfunc}
\end{align}
where $K(Q;\alpha,R)$ is the so-called Coulomb correction factor (calculated numerically), and the $N(1+\varepsilon Q)$ factor represents a possible linear background (usually negligible).
\section{Data analysis}

In this section, we detail the process of the data analysis at STAR, in particular, the measurement and fitting of the two-particle correlation functions. The data set used for the analysis was recorded in 2011 by the STAR experiment, in Au+Au collisions at $\sqrt{s_{NN}} = 200$~GeV center-of-mass collision energy. The minimum-bias data contained about 550 million events. 

The main detector of the STAR experiment is a Time Projection Chamber (TPC), used for centrality determination, vertex position measurement, tracking, and particle identification with ionization energy loss (dE/dx)~\cite{Anderson:2003ur}. Another important detector used in this analysis is the barrel Time Of Flight (bTOF) detector~\cite{Shao:2005iu,Wu:2007zzk, Llope:2011zz}, which is used together with the TPC to cut out pile-up events from the sample and also to aid particle identification, especially at higher transverse momentum ranges.

After a careful event selection based on vertex position cuts and pile-up cuts, the next step was to select the charged pion tracks. The track selection criteria included cuts on the number of hits used to reconstruct the track in the TPC ($N_{\rm hits} > 20$), on the distance of closest approach to the primary vertex (DCA < 2.0 cm), on pseudo-rapidity ($|\eta| < 0.75$), and on transverse momentum ($0.15 < p_T [{\rm GeV}/c] < 1.0$). For the pion identification, a combined approach was utilized using the dE/dx measured by TPC and the time-of-flight measured by TOF simultaneously. We found that this eliminates the need for veto cuts; however, when only TPC information was available, veto cuts for electrons, kaons, and protons were also utilized. In measurements where quantities such as relative and average momentum are calculated for pairs of particles, another important aspect is taking into account the merging and splitting effects stemming from the track reconstruction algorithm~\cite{STAR:2004qya}. For the latter, a cut on the splitting level quantity was used (SL < 0.6), similarly to Ref.~\cite{STAR:2004qya}. To correct for the merging effect, a cut on the fraction of merged hits (FMH) quantity was used (similarly to Ref.~\cite{STAR:2004qya}), requiring it to be less than 5 percent. Furthermore, the average separation of the pair was calculated over the TPC pad rows and was required to be greater than 3 cm.

To construct the correlation functions, the event-mixing method was utilized, as described in, e.g., Ref.~\cite{PHENIX:2017ino}. When mixing pairs from separate events, both events were required to belong to the same event class, using 2 cm $z$-vertex bins and 5\% centrality bins. Correlation functions were constructed for four centrality bins (0-10\%, 10-20\%, 20-30\%, 30-40\%), and for 21 average transverse momentum $k_T$ bins, ranging from 0.175 GeV/$c$ up to 0.750 GeV/$c$. The one-dimensional relative-momentum variable of choice was the magnitude of the three-momentum difference in the longitudinal co-moving system (LCMS)~\cite{Kurgyis:2020vbz}.

\begin{figure}[H]
\includegraphics[width=\textwidth]{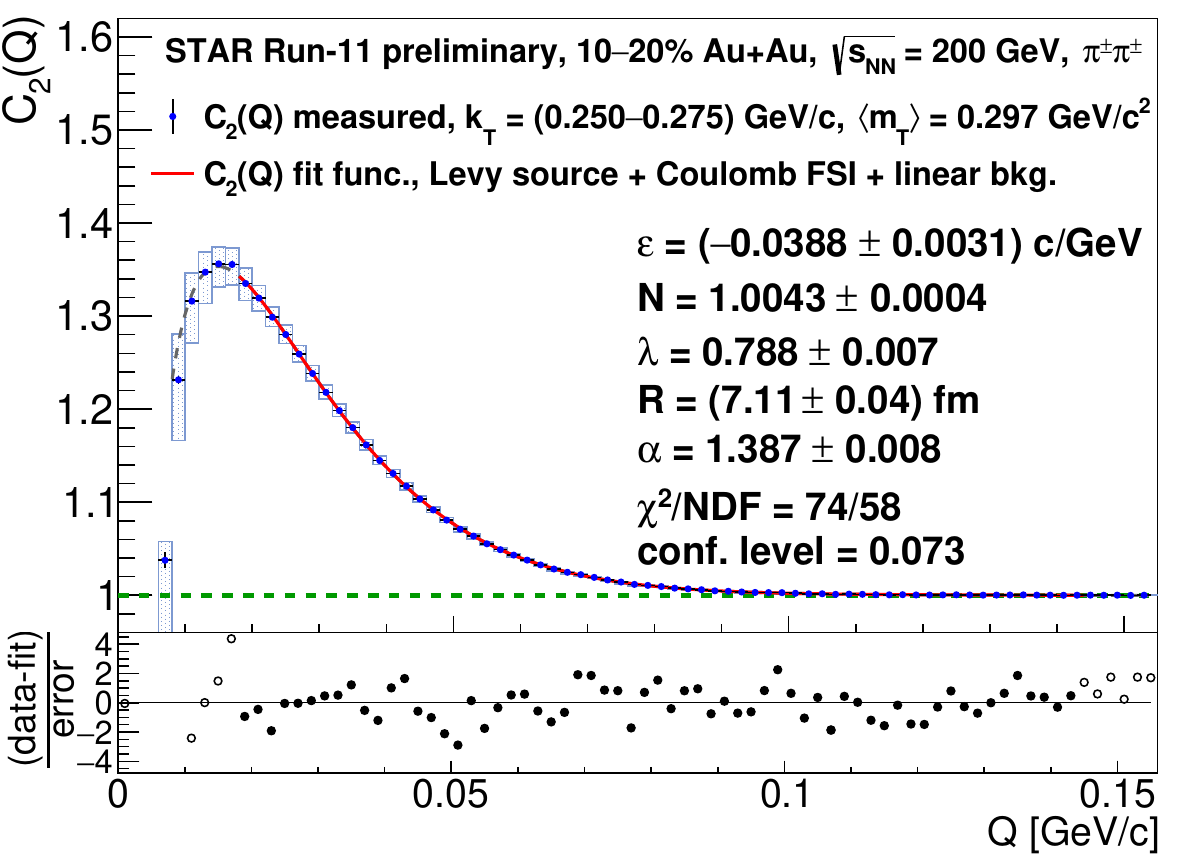}
\caption{An example two-pion correlation function measured in $\sqrt{s_{NN}}$~=~200~GeV Au+Au collisions. The correlation function belongs to the 10-20\% centrality class, with pion pairs having an average transverse momentum in the range of 0.250 GeV/$c$--0.275 GeV/$c$. The measured correlation function is denoted by dark blue data points with statistical uncertainties marked by error bars. The opaque blue boxes on the data points correspond to the systematic uncertainties stemming from single-track and pair cuts and variations. The fit function (corresponding to Eq.~\eqref{e:fitfunc}) is shown with a red curve within the fit range and with a grey dashed curve outside.  \label{f:fitexample}}
\end{figure}

Fitting of the correlation functions was done using Eq.~\eqref{e:fitfunc} in an iterative, self-consistent way, similarly to what is described in Ref.~\cite{PHENIX:2017ino}. An example fit is shown in Fig.~\ref{f:fitexample}. Similarly to Fig.~\ref{f:fitexample}, all other fits also converged with a confidence level greater than 0.1\%. The systematic uncertainty investigations included variations of the previously detailed single-track and pair cuts and variations of the lower and upper fit limits. The effect of the choice of bin width in the relative momentum variable and the difference between $\pi^+\pi^+$ and $\pi^-\pi^-$ correlation functions were found to be negligible.

\section{Results}

In this section, we review the centrality and the average transverse mass ($m_T$) dependence of the extracted source parameters.

\begin{figure}[H]
\includegraphics[width=\textwidth]{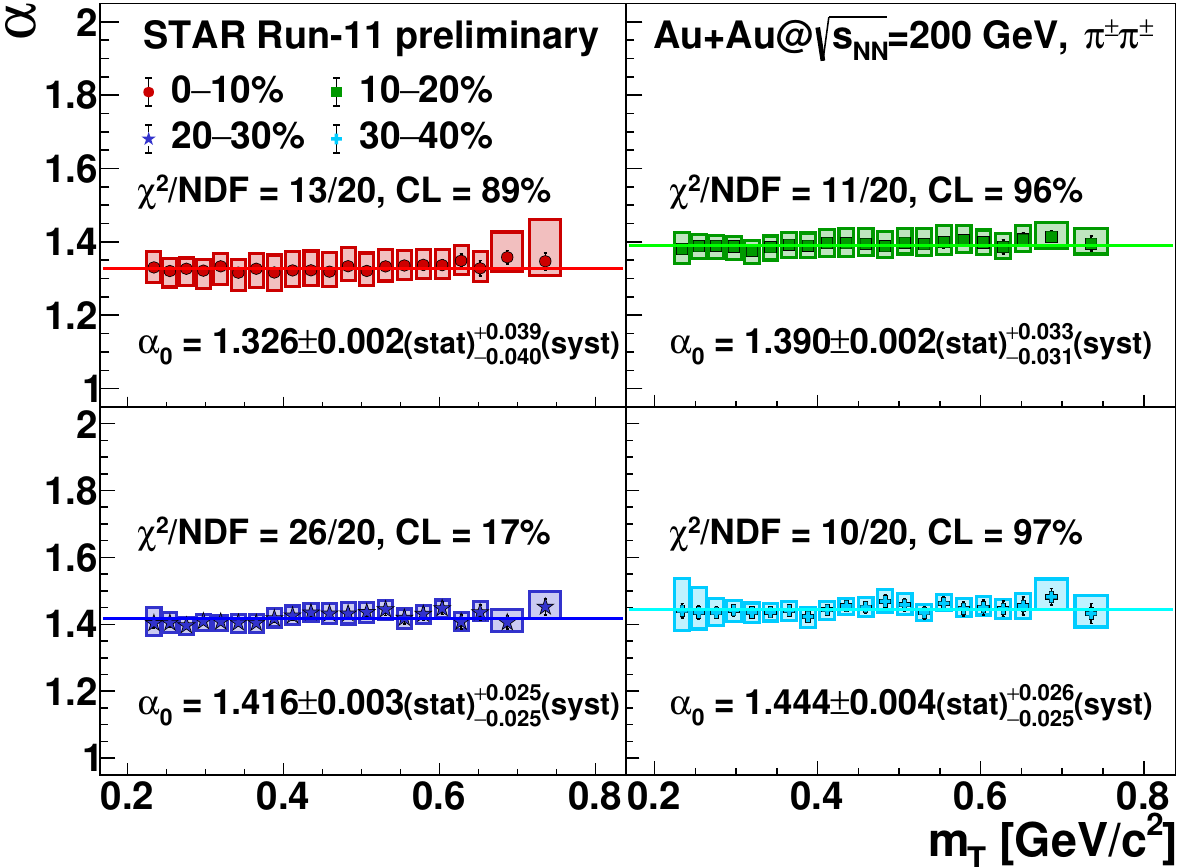}
\caption{Average transverse momentum dependence of the Lévy exponent $\alpha$ parameter in Au+Au collisions at $\sqrt{s_{NN}}$~=200~GeV for four different centrality classes (0--10\%, 10--20\%, 20--30\%, 30--40\%). The $m_T$ dependence is fitted with a constant $\alpha(m_T) = \alpha_0$ parametrization, which provides a statistically acceptable description at all four centrality classes. The colored markers and error bars correspond to the parameter values and their statistical uncertainties extracted from fits similar to Fig.~\ref{f:fitexample}. The colored boxes correspond to the total systematic uncertainties. \label{f:alphavsmt}}
\end{figure}

\begin{figure}[H]\centerline{
\includegraphics[width=0.75\textwidth]{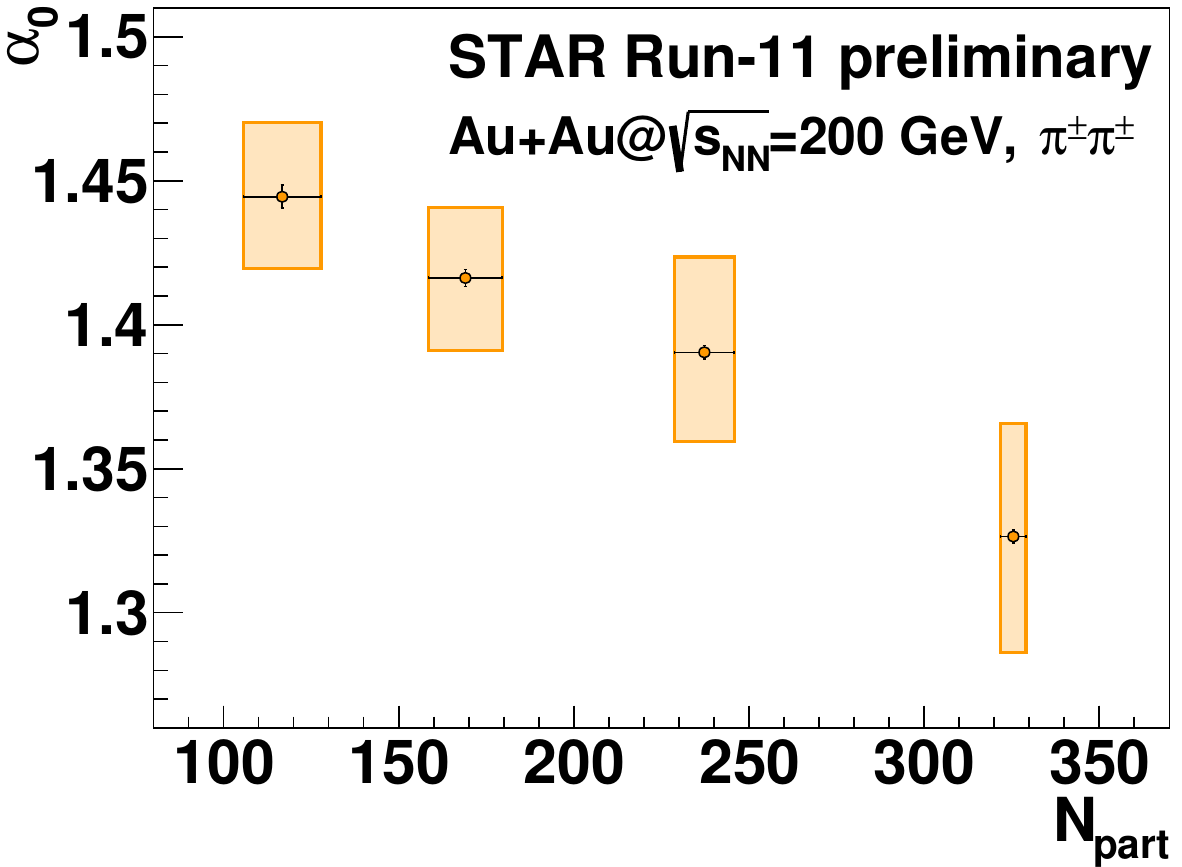}}
\caption{Number of participant nuclei dependence of the $m_T$ averaged L\'evy exponent. The colored markers and error bars correspond to the parameter values and their statistical uncertainties extracted from the $\alpha(m_T) = \alpha_0$ constant fits. The colored boxes correspond to the total systematic uncertainties.\label{f:alpha0vsnpart}}
\end{figure}

The L\'evy exponent $\alpha$ as a function of $m_T$ and centrality is shown in Fig.~\ref{f:alphavsmt}. There is very little dependence on the average transverse mass; a constant fit provides a good description of $\alpha(m_T)$ at all four centrality classes. The values are far from the Gaussian ($\alpha=2$) case and decrease with increasing $N_{\rm part}$ values, as shown in Fig.~\ref{f:alpha0vsnpart}. It is interesting to note that the CMS experiment observed an opposite trend with centrality~\cite{CMS:2023xyd}, albeit with different kinematic cuts and no particle identification; hence, a direct comparison is difficult. As $\alpha$ is highly anti-correlated with the other two fit parameters, this could be mirrored in the $N_{\rm part}$ dependence as well; as $N_{\rm part}$ (and the multiplicity) increases, the size of the system (and with that the $R$ L\'evy scale parameter) increases, and $\alpha$ decreases.

\begin{figure}[H]\centerline{
\includegraphics[width=\textwidth]{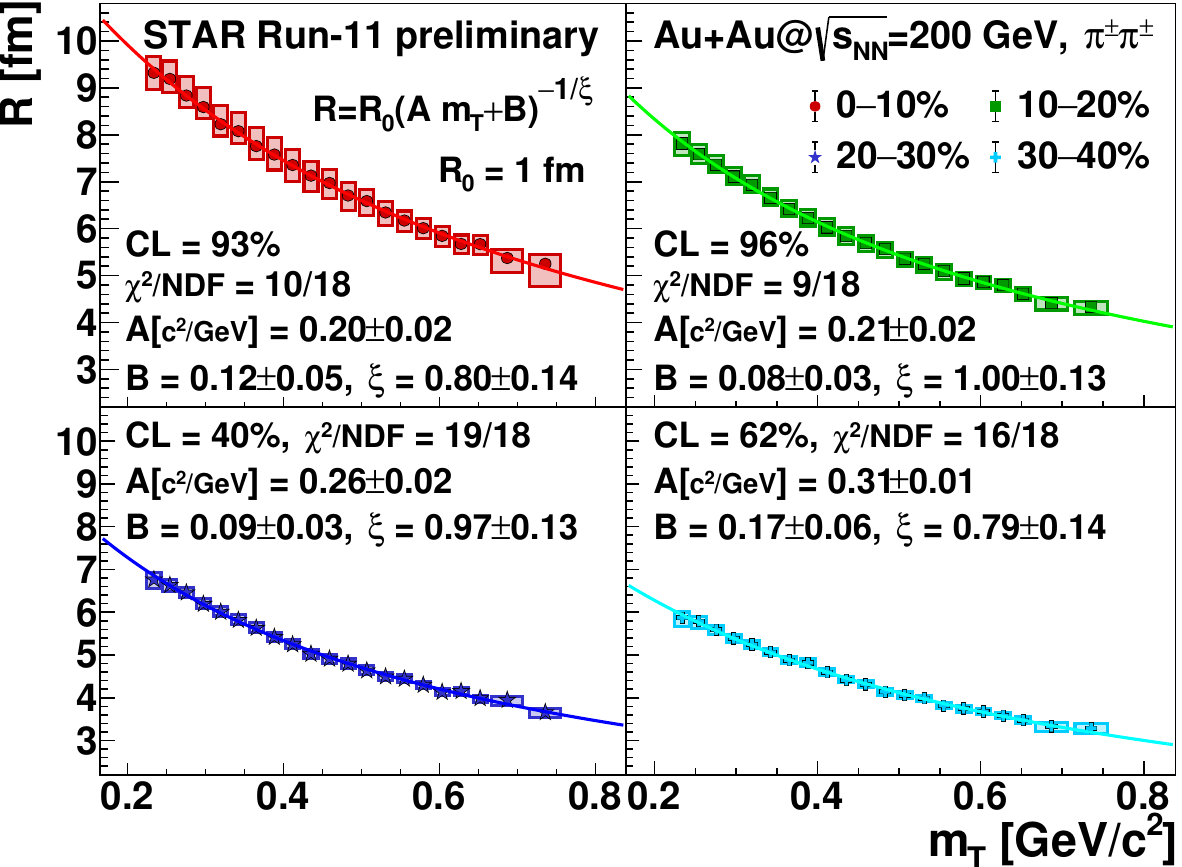}}
\caption{Average transverse momentum dependence of the Lévy scale $R$ parameter in Au+Au collisions at $\sqrt{s_{NN}}$~=200~GeV for four different centrality classes (0--10\%, 10--20\%, 20--30\%, 30--40\%), plotted on separate panels. The $m_T$ dependence is fitted with an ${R(m_T) = R_0(Am_T+B)^{-1/\xi}}$ parametrization, which provides a statistically acceptable description at all four centrality classes. The colored markers and error bars correspond to the parameter values and their statistical uncertainties extracted from fits similar to Fig.~\ref{f:fitexample}. The colored boxes correspond to the total systematic uncertainties. \label{f:rvsmt_multi}}
\end{figure}

The L\'evy scale parameter $R$ is shown in Fig.~\ref{f:rvsmt_multi} with the four centrality classes on separate panels. The decreasing trend with $m_T$ is very similar to the usual observations for Gaussian HBT radii~\cite{STAR:2014shf} and might be attributed to the hydrodynamic expansion of the system~\cite{Makhlin:1987gm, Csorgo:1995bi, Lisa:2008gf}. Hydrodynamic calculations for Gaussian radii predict an $R\propto 1/\sqrt{m_T}$ type scaling. The PHENIX experiment found that this scaling holds for the L\'evy-scale parameter as well~\cite{PHENIX:2017ino}, however, with the level of precision available at the STAR experiment, this might not hold anymore. Fig.~\ref{f:rvsmt_multi} includes fits to the transverse mass dependence parameterized as ${R(m_T) = R_0(Am_T+B)^{-1/\xi}}$. The fits provide a statistically acceptable description at all centrality classes and the values of the $\xi$ exponent are not compatible with $\xi=2$ (which would correspond to the hydro calculations). These interesting observations provide ample motivation for new theoretical and phenomenological studies involving hydrodynamical calculations with L\'evy-type sources.

\begin{figure}[H]
\centerline{\includegraphics[width=\textwidth]{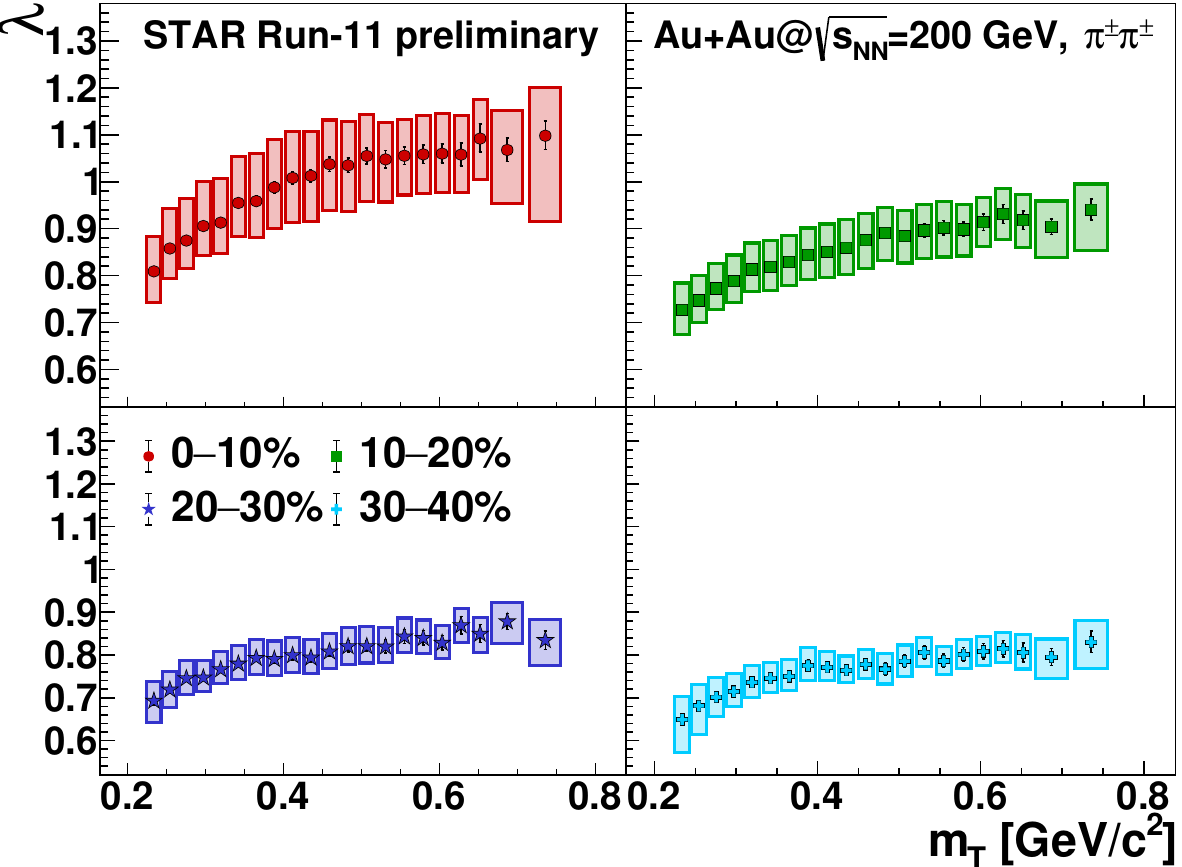}}
\caption{Average transverse momentum dependence of the correlation strength $\lambda$ parameter in Au+Au collisions at $\sqrt{s_{NN}}$~=200~GeV for four different centrality classes (0--10\%, 10--20\%, 20--30\%, 30--40\%). The colored markers and error bars correspond to the parameter values and their statistical uncertainties extracted from fits similar to Fig.~\ref{f:fitexample}. The colored boxes correspond to the total systematic uncertainties. \label{f:lambdavsmt}}
\end{figure}

Last but not least, the correlation strength parameter $\lambda$ is shown in Fig.~\ref{f:lambdavsmt}. It exhibits very similar behavior with $m_T$ as the PHENIX result of Ref.~\cite{PHENIX:2017ino}, there is an increase and a saturation towards high-$m_T$. The magnitude of the parameter also depends on centrality; the highest values are observed in the most central case. There can be many reasons behind the observed $m_T$ dependence, e.g., in-medium mass modification of the $\eta'$ meson~\cite{PHENIX:2017ino}, or partial coherence~\cite{Csorgo:1999sj}. Within the core-halo model~\cite{Csorgo:1994in}, the $\lambda$ parameter is interpreted as the squared fraction of primordial pions (and decays of short-lived resonances) to all produced pions (including decays of long-lived resonances). This is, however, not yet well-understood for power-law sources and needs to be explored in more detail from the phenomenology side. It is also important to note that a different trend was observed at SPS energies with no decrease at low $m_T$~\cite{NA61SHINE:2023qzr,Porfy:2023yii}. Hence, a beam energy dependent analysis could provide interesting new insights into the interpretation of this parameter as well.





\section{Discussion, conclusions, and outlook}

In this paper, we presented comprehensive new preliminary results on the average transverse mass and centrality dependence of the L\'evy source parameters of pion pairs. The source parameters $\alpha$, $R$, and $\lambda$ were extracted from one-dimensional momentum correlation functions measured by the STAR experiment in $\sqrt{s_{NN}}$~=~200~GeV Au+Au collisions. 

The values of the L\'evy exponent $\alpha$ were in the range of 1.3-1.5, far from the Gaussian ($\alpha = 2$) case. No $m_T$ dependence and a decreasing trend from peripheral to central collisions were observed. The L\'evy scale $R$ parameter showed a clear centrality ordering, possibly connected to initial geometry, with the largest values observed at the most central case. A decreasing trend with $m_T$ was also observed, similar to observations for the Gaussian HBT radii. The $m_T$ dependence was fitted with an ${R(m_T) = R_0(Am_T+B)^{-1/\xi}}$ parametrization, which provided a good description at all four centrality classes. The correlation strength parameter showed an increasing trend and a saturation at high-$m_T$ values, and a decreasing trend from central to peripheral collisions. The trends and magnitudes of the parameters are also quite close to published PHENIX results for 0-30\% Au+Au collisions at the same collision energy~\cite{PHENIX:2017ino}.

To finalize these results, a more detailed systematic uncertainty investigation is currently underway. Furthermore, this L\'evy analysis is being extended to lower collision energies recorded during the second phase of the RHIC Beam Energy Scan. Ongoing and planned investigations also include a three-dimensional analysis of the pion correlation functions and a kaon analysis. These new experimental results, together with phenomenological investigations, will hopefully shed light on the physical processes playing a role in shaping the two-particle source function in heavy-ion collisions. 


\funding{This research was funded in part by the NKFIH grants K-138136, TKP2021-NKTA-64, and PD-146589.}

\dataavailability{Data sharing is not applicable to this article.} 


\conflictsofinterest{The authors declare no conflict of interest.} 




\begin{adjustwidth}{-\extralength}{0cm}

\reftitle{References}



\end{adjustwidth}
\end{document}